\documentclass[5p,twocolumn]{elsarticle}
\usepackage[utf8]{inputenc}
\usepackage[T1]{fontenc}
\usepackage{xcolor}
\usepackage[hidelinks]{hyperref}
\hypersetup{
  pdftitle={-},
  pdfauthor={-}
}
\usepackage{graphicx}
\usepackage{amsmath}
\usepackage{cases}
\usepackage{dsfont}
\usepackage{mathrsfs}
\usepackage{braket}
\usepackage{blkarray}
\usepackage[integrals]{wasysym}
\usepackage{cuted}
\usepackage{caption}
\usepackage{mathtools}
\usepackage{dcolumn}
\usepackage{siunitx}
\usepackage[normalem]{ulem}
\usepackage{multirow}
\usepackage{booktabs}
\usepackage{tablefootnote}
\newcommand{\threej}[6]{
    \begin{pmatrix}
    #1 & #2 & #3 \\
    #4 & #5 & #6
    \end{pmatrix}
}
\newcommand{\sixj}[6]{
    \begin{Bmatrix}
    #1 & #2 & #3 \\
    #4 & #5 & #6
    \end{Bmatrix}
}
\newcommand{\wigthree}{
    \big(
    \begin{smallmatrix}
    . & . & .\\ 
    . & . & .
    \end{smallmatrix}
    \big)
}
\newcommand{\wigsix}{
    \big\lbrace
    \begin{smallmatrix}
    . & . & .\\ 
    . & . & .
    \end{smallmatrix}
    \big\rbrace
}
\biboptions{sort&compress}
\begin{document}

\begin{frontmatter}
\title{Intensities of all fine-structure resolved rovibrational electric quadrupole absorption lines in $^{16}$O$_2$($X^{3}\Sigma^{-}_{g}$) calculated with a new \textit{ab initio} quadrupole moment curve}

\author[1]{Maciej Gancewski\corref{corr1}}
\ead{ganc@doktorant.umk.pl}
\author[1]{Hubert Jóźwiak}
\author[2]{Hubert Cybulski}
\author[1]{Piotr Wcisło}
\ead{piotr.wcislo@umk.pl}

\cortext[corr1]{Corresponding author}

\affiliation[1]{
    organization = {Institute of Physics, Faculty of Physics, Astronomy and Informatics, Nicolaus Copernicus University in Toruń},
    street = {Grudziądzka 5},
    city         = {Toruń},
    postcode     = {87-100},
    country      = {Poland}}
\affiliation[2]{
    organization = {Faculty of Physics, Kazimierz Wielki University},
    street = {al. Powstańców Wielkopolskich 2,},
    city         = {Bydgoszcz},
    postcode     = {85-090},
    country      = {Poland}}

\begin{abstract}
The intensities of all rovibrational electric quadrupole absorption lines in $^{16}$O$_2$($X^{3}\Sigma^{-}_{g}$), for which the vibrational quantum number is $v \leq 35$ and the total angular momentum quantum number is $J \leq 40$, are calculated in the intermediate coupling using a new \textit{ab initio} quadrupole moment curve of the ground electronic state of O$_2$. The calculated values agree with those available in the HITRAN database, which at present includes only the $1$-$0$ fundamental vibrational band of $^{16}$O$_2$($X^{3}\Sigma^{-}_{g}$). We therefore recommend using the intensities of the vibrational overtones and hot bands reported here in updating the HITRAN database for O$_2$ in the upcoming 2024 edition.
\end{abstract}


\end{frontmatter}

\section{Introduction}
In 1980, Niple \textit{et al.}~\cite{niple1980} published stratospheric spectra obtained with a balloon-borne interferometer and used them to identify solar features of various atmospheric molecules. In assigning the relevant spectra, however, they encountered a peculiar, closely spaced triplet of lines near $\sim$1604 cm$^{-1}$, which was left unidentified in their work. A year later, Goldman, Reid and Rothman~\cite{goldman1981} suggested that these lines were due to the fine-structure resolved $S(7)$ electric quadrupole (\textit{E2}) transition in the $1$-$0$ fundamental vibrational band of $^{16}\text{O}_2$ in its ground electronic $X^3\Sigma^-_g$ term. They supported this claim by calculating relevant transition frequencies and comparing them with those obtained in previous lower-resolution measurements of the rovibrational Raman spectra of O$_2$~\cite{fletcher1974, loete1977}, and by showing that the strongest calculated linestrength in the band corresponds to the aforementioned $S(7)$ line, making its observation under atmospheric conditions highly plausible. In their short letter~\cite{goldman1981}, they further announced a future publication of the relevant experimental data confirming their findings. 

In the meantime, Rothman and Goldman~\cite{rothman1981} calculated, for the first time, the intensities of all fine-structure resolved \textit{E2} transitions in the $1$-$0$ fundamental band of $^{16}\text{O}_2(X^3\Sigma^-_g)$ that were allowed by the selection rules of the Hund's case ($b$) approximation, which they employed for the description of the triplet fine structure of individual rotational energy levels in O$_2$. The key deficiency of this calculation was the lack of an explicit \textit{E2} transition moment (to which squared modulus the line intensity is directly proportional). The employed formula for the line intensities contained instead a free parameter adjusted to the soon-to-be-published laboratory measurements announced earlier in Ref.~\cite{goldman1981}. For this reason, the confidence in the calculated line intensities was assumed to be $\pm 35\%$, the same as the estimated uncertainty of the measurements.  

Later the same year, the laboratory measurements of Reid \textit{et al.}~\cite{reid1981} concerning the intensities of \textit{E2} lines in the $1$-$0$ band of O$_2$ were finally published. This was the first observation of a vibrational \textit{E2} transition in a molecule other than hydrogen. These measurements decidedly confirmed that the unidentified triplet of lines encountered in the solar spectra by Niple \textit{et al.}~\cite{niple1980} was due to the $J = 8 \leftarrow 6$, $9 \leftarrow 7$ and $10 \leftarrow 8$ fine structure components of the rovibrational $1$-$0$ $S(7)$ transition in the ground-electronic state of O$_2$. Reid \textit{et al.}~\cite{reid1981} managed to observe also the $J = 6 \leftarrow 4$, $7 \leftarrow 5$ and $8 \leftarrow 6$ components of the $1$-$0$ $S(5)$ rovibrational line, albeit with a slightly lower accuracy due to the relative weakness of this transition. No other rovibrational \textit{E2} lines were measured due to the insufficient sensitivity of the instruments. In view of this, it is worth noting that all the observed fine structure components correspond to the $\Delta J = \pm 2$ selection rules, i.e., they are purely \textit{E2} transitions. There are, however, also the $\Delta J = 0, \pm 1$ selection rules, which are common to both the \textit{E2} and \textit{M1} (magnetic dipole) transitions, and it seems interesting to investigate the relative contribution of these two mechanisms to the total spectral line intensity. Thus, Reid \textit{et al.}~\cite{reid1981} tuned their laser to look for the $J = 8 \leftarrow 7$ fine structure component of the $S(7)$ line, which in such a case would be one of the strongest \textit{M1} lines in the $1$-$0$ band. Curiously, however, they did not detect any absorption in this spectral region, thus being only able to put an upper limit on the line-center absorption coefficient. Nonetheless, they used the available numerical routines~\cite{lepard1970, johns1975} to estimate the relative contributions to the overall $1$-$0$ band intensity due to the observed six fine-structure components of the $S(5)$ and $S(7)$ \textit{E2} lines as well as due to the unobserved components of these and other \textit{E2} transitions; they further estimated the analogous relative intensities of \textit{M1} transitions, which could not be detected in their experiment.

In 1990, Balasubramanian, D'Cunha and Rao~\cite{balasubramanian1990}, building on Rothman's and Goldman's work~\cite{rothman1981}, recalculated the intensities of all fine-structure resolved \textit{E2} transitions in the $1$-$0$ fundamental band of $^{16}\text{O}_2(X^3\Sigma^-_g)$ using the intermediate coupling, i.e., going beyond the Hund's case ($b$) approximation employed in Ref.~\cite{rothman1981}. Using the exact eigenvectors of the effective Hamiltonian in the line-intensity calculations allowed them to account for additional contributions to transition intensities due to mixing of rotational states of O$_2$ induced by the relativistic spin-spin interaction (e.g., pure Hund's case ($b$) overestimates the intensity of the $J = 2 \leftarrow 1$ component of the $1$-$0$ $Q(1)$ line by $\sim$30\%~\cite{balasubramanian1990}) and resulted further in the occurrence of two ``extra'' rotational branches, corresponding to the apparent selection rules $\Delta N = -4$ ($M$-branch) and $\Delta N = +4$ ($U$-branch) for the rotational quantum number $N$. Despite this refinement of Rothman's and Goldman's results~\cite{rothman1981}, the expression for the line intensity employed by Balasubramanian, D'Cunha and Rao~\cite{balasubramanian1990} also contained a free parameter, rather than the explicit \textit{E2} transition moment, which was fitted to the laboratory measurements of Reid \textit{et al.}~\cite{reid1981} so as to match their results for the $J = 11 \leftarrow 9$ component of the $1$-$0$ $S(9)$ line. This line, however, was not actually measured by Reid \textit{et al.}~\cite{reid1981} who, as was already mentioned, only reported its relative contribution to the overall \textit{E2} $1$-$0$ band intensity based on numerical estimates. A deficiency on the theory's side therefore still remained.

That same year, Dang-Nhu \textit{et al.}~\cite{dang-nhu1990} identified first rovibrational \textit{M1} lines in the $1$-$0$ fundamental band of O$_2$, which Reid \textit{et al.}~\cite{reid1981} previously failed to observe, based on atmospheric spectra obtained by the ATMOS experiment~\cite{farmer1987} and the (unpublished) University of Denver balloon-borne stratospheric scans. In their work~\cite{dang-nhu1990}, relevant transition frequencies were recalculated using the intermediate coupling, whereas the line intensities themselves were obtained in terms of a rather questionable phenomenological expression,\footnote{\label{foot1}The mechanism governing the rovibrational \textit{M1} transitions was explained in Ref.~\cite{dang-nhu1990} on the basis of a formal analogy with the electric dipole transitions. This point was raised and criticized by Balasubramanian, Bellary and Rao~\cite{balasubramanian1994} who proposed that the \textit{M1} components of the $1$-$0$ oxygen band arise due to spin-orbit mixing, and recalculated both the intensities and the corrections due to possible rotation-vibration interaction.} involving band intensity as a free parameter appropriately fitted to the few identified \textit{M1} lines; the relative \textit{M1} intensities thus obtained differed considerably from the corresponding numerical estimates made previously by Reid \textit{et al.}~\cite{reid1981}. The identification of the first rovibrational \textit{M1} and \textit{E2} oxygen transitions in the atmospheric spectra and the availability of the newly generated spectroscopic datasets (though based on empirical fits, rather than \textit{ab initio} multipole moments) presented a possibility for the mutual refinement of the available \textit{M1} and \textit{E2} data and the subsequent accurate modeling of the entire $1$-$0$ fundamental band of O$_2$. 

The need for high-quality datasets for the $1$-$0$ band required for applications prompted additional investigations. In 1992, Balasubramanian \textit{et al.}~\cite{balasubramanian1992} determined the corrections to the previously obtained \textit{E2} intensities~\cite{balasubramanian1990} induced by the rotation-vibration interaction. In 1994, Balasubramanian, Bellary and Rao~\cite{balasubramanian1994} recalculated the \textit{M1} intensities (cf. footnote~\ref{foot1}) and determined the possible rotation-vibration interaction-induced corrections to these transitions. In 1995, Goldman \textit{et al.}~\cite{goldman1995} discussed the available data on the oxygen $1$-$0$ band with the prospect of updating the HITRAN spectroscopic database~\cite{hitran2022}. In this work~\cite{goldman1995}, they recalculated the \textit{M1} intensities for different parallel-to-perpendicular ratios of the components of the magnetic dipole moment and further recommended that the intermediate-coupling \textit{E2} intensities obtained by Balasubramanian, D'Cunha and Rao~\cite{balasubramanian1990} be used, for consistency, in the preparation of relevant datasets for the oxygen $1$-$0$ band. When compiling the available data on the isotopologues of O$_2$ in 1998, however, Gamache, Goldman and Rothman~\cite{gamache1998} determined the intensities of \textit{E2} transitions in the $1$-$0$ band of $^{16}\text{O}_2(X^3\Sigma^-_g)$ by scaling the relative intensities (using the appropriate factors that were determined in Ref.~\cite{goldman1995}) which were estimated numerically by Reid \textit{et al.}~\cite{reid1981}. A slight confusion arose~\cite{gamache1998}, when comparing these intensities with the previous 1992 HITRAN edition, as a result of incorrect identification of the \textit{E2} transition moment with the vibrationally-averaged ``permanent'' \textit{E2} moment of the ground-state O$_2$. The current 2020 edition of the HITRAN database~\cite{hitran2022} lists the intensities of the rovibrational \textit{E2} transitions in $^{16}\text{O}_2(X^3\Sigma^-_g)$ only for the case of the $1$-$0$ fundamental band; at the time of writing this article, the HITRAN website~\cite{hitran_website} cites Ref.~\cite{gamache1998} as the source of these intensities.

Despite the fact that the theoretical investigations described above did not make use of precalculated \textit{ab initio} transition moments in obtaining the rovibrational \textit{E2} line intensities, a body of work devoted to the determination of the \textit{E2} moments of $^{16}\text{O}_2(X^3\Sigma^-_g)$ have appeared throughout the years. In particular, the vibrationally-averaged ``permanent'' \textit{E2} moment of the ground-state O$_2$ was investigated both empirically~\cite{buckingham1968, birnbaum1976, cohen1977, couling2014} as well as \textit{ab initio}~\cite{becke1982, vanLenthe1984, visser1985, laaksonen1985, rijks1989, hettema1994, bundgen1995, dickinson1996, lawson1997, minaev2004, bartolomei2010, kalugina2015, somogyi2024} (see Tab.~\ref{tab: quadrupole moments 1} in Sec.~\ref{Sec3} below), though it seems that the transition \textit{E2} moment (corresponding to the $1$-$0$ fundamental band) was determined only empirically~\cite{reid1981}.

In this work, we calculate the intensities of all rovibrational \textit{E2} absorption lines in the ground-electronic $^{16}\text{O}_2$ isotopologue (including vibrational overtones and hot bands) using a new \textit{ab initio} \textit{E2} moment curve of $^{16}\text{O}_2(X^3\Sigma^-_g)$. We account for the spin-spin interaction-induced mixing of rotational levels of O$_2$ and perform all calculations in the intermediate coupling, i.e., we go beyond the Hund's case ($b$) approximation. Our \textit{ab initio} results for the $1$-$0$ fundamental band differ from the values presently available in HITRAN~\cite{hitran2022} at the $5\%-12\%$ level, depending on the fine-structure resolved rovibrational line. The uncertainties of the HITRAN2020 line intensities for this band are listed as ``unreported or unavailable''~\cite{hitran_website}. In view of the above, we recommend incorporating the intensities of vibrational overtones and hot bands reported here in the new edition of the HITRAN database.

\section{General equations}
\subsection{Spin-spin interaction-induced mixing of rotational states of O$_2$ in the intermediate coupling}
The theoretical description of the ground $^3\Sigma^-_g$ electronic term of O$_2$ and the triplet fine structure of its rotational energy levels was given and refined by many authors~\cite{vanVleck1928, kramers1929, hebb1936, schlapp1937, miller1953, mizushima1954, tinkham1955, steinbach1973}. In this section, we review the basic facts on the relativistic spin-spin interaction-induced mixing of rotational states of O$_2$ responsible for the breakdown of the Hund's case ($b$) picture and we set up our notation and conventions.

The effective Hamiltonian $\mathcal{H}^{(\text{eff})}_v$ of $^{16}\text{O}_2(X^3\Sigma^-_g)$ in a vibrational state $v$ can be written as~\cite{Gordy&Cook, yu2012}
\begin{align}
    \mathcal{H}^{(\text{eff})}_v = G^{(\text{vib})}_v + A^{(\text{rot})}_v\mathbf{N}^2 + A^{(\text{sr})}_v\mathbf{N}\cdot\mathbf{S} + A^{(\text{ss})}_v\frac{2}{3}\left(3S_z^2 - \mathbf{S}^2\right) \rm{,}
    \label{eq: H_eff}
\end{align}
where $G^{(\text{vib})}_v$ is the vibrational energy and the three subsequent terms are the rotational (rot) Hamiltonian, the spin-rotation (sr) interaction and the spin-spin (ss) interaction, respectively. $\mathbf{N}$ denotes the sum of the end-over-end rotational angular momentum and the total electronic orbital angular momentum, and $\mathbf{S}$ and $S_z$ are the operators of the total electronic spin and its component along the space-fixed $z$-axis, respectively. The corrections due to centrifugal distortion are accounted for by expanding the effective molecular constants $A^{(i)}_v$ in Eq.~\eqref{eq: H_eff} in powers of $\mathbf{N}^2$. In this work we use~\cite{yu2012}
\begin{subequations}\label{eqs: effective constants}
\begin{align}
    &A^{(\text{rot})}_v = B_v + D_v\mathbf{N}^2 + H_v\left(\mathbf{N}^2\right)^2 \rm{,} \\
    &A^{(\text{sr})}_v = \gamma_v + \gamma_{D_v}\mathbf{N}^2 + \gamma_{H_v}\left(\mathbf{N}^2\right)^2 \rm{,} \\
    &A^{(\text{ss})}_v = \lambda_v + \lambda_{D_v}\mathbf{N}^2 + \lambda_{H_v}\left(\mathbf{N}^2\right)^2 + \lambda_{L_v}\left(\mathbf{N}^2\right)^3 \rm{,}
\end{align}
\end{subequations}
where the expansion coefficients \big(as well as $G^{(\text{vib})}_v$ in Eq.~\eqref{eq: H_eff}\big) are those of Yu \textit{et al.}~\cite{yu2014}.

Matrix elements of $\mathcal{H}^{(\text{eff})}_v$ are most easily calculated using the Hund's case ($b$) basis, i.e., the basis of eigenfunctions $\ket{(NS)J}=\ket{(NS)JM}$ of the total angular momentum operator $\mathbf{J}=\mathbf{N}+\mathbf{S}$ (for the time being, the space-fixed projection quantum number $M$ can be omitted from the state vector). The rotational and spin-rotation terms in Eq.~\eqref{eq: H_eff}, as well as the expansions in Eqs.~\eqref{eqs: effective constants}, are diagonal in this basis, while the matrix elements of the spin-spin interaction term can be calculated using the formula~\cite{judd1975}
\begin{align}
    &\braket{(NS)J|(3S_z^2-\mathbf{S}^2)|(N'S')J'} = \delta_{SS'}\delta_{JJ'} \times \notag\\
    &\times (-1)^{J+1}\sqrt{30}[N,N']^\frac{1}{2} \threej{N}{2}{N'}{0}{0}{0} \sixj{N}{2}{N'}{1}{J}{1} \rm{,}
    \label{eq: <spin-spin>}
\end{align}
where $[a,b] = (2a+1)(2b+1)$ is introduced as a shorthand notation, while $\wigthree$ and $\wigsix$ are the $3j$ and $6j$ Wigner symbols, respectively \big(note that the total electronic spin quantum number $S=1$ is included explicitly in Eq.~\eqref{eq: <spin-spin>}\big).  

\begin{figure}[h!]
    \centering
        \includegraphics[width=\linewidth]{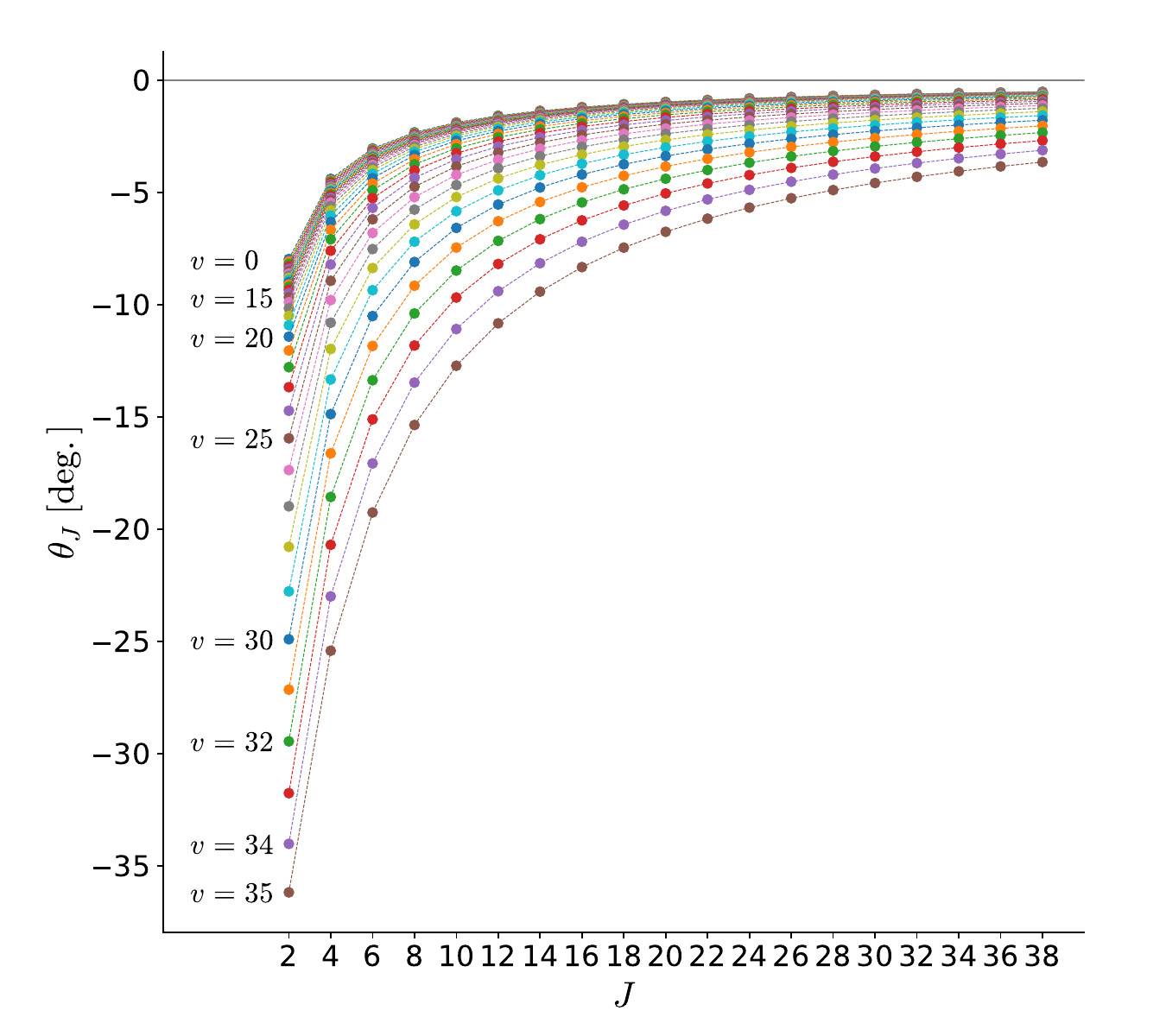}
    \caption{Mixing angles (in degrees) calculated using Eq.~\eqref{eq: mixing angle} for all vibrational ($v$) and total angular momentum ($J$) quantum numbers considered in this work (for clarity, only some values of $v$ are indicated explicitly). Note that the mixing occurs only for even values of $J \neq 0$.}
    \label{fig: mixing angle}
\end{figure}

Equation~\eqref{eq: <spin-spin>} shows that the spin-spin interaction mixes the pairs of Hund's case ($b$) states for which $\Delta N = \pm 2$. Since in $^{16}\text{O}_2(X^3\Sigma^-_g)$ the rotational quantum number $N$  can only take \textit{odd} values (due to nuclear spin statistics), the matrix of $\mathcal{H}^{(\text{eff})}_v$ (for a given value of $v$) written in the $\ket{N, J}=\ket{(NS)J}$ basis has the form
\begin{equation}
\makeatletter\setlength\BA@colsep{2.0pt}\makeatother
\begin{blockarray}{rccccccccc}
              & & \ket{1,0} & \ket{1,1} & \ket{1,2} & \ket{3,2} & \ket{3,3} & \ket{3,4} & \ket{5,4} & \dots \\
    \begin{block}{cc(cccccccc)}
    \ket{1,0} & & h^{(0)}_{11} & 0 & 0 & 0 & 0 & 0 & 0 & \dots \\
    \ket{1,1} & & 0 & h^{(1)}_{11} & 0 & 0 & 0 & 0 & 0 & \dots \\
    \ket{1,2} & & 0 & 0 & h^{(2)}_{11} & h^{(2)}_{13} & 0 & 0 & 0 & \dots \\
    \ket{3,2} & & 0 & 0 & h^{(2)}_{31} & h^{(2)}_{33} & 0 & 0 & 0 & \dots \\
    \ket{3,3} & & 0 & 0 & 0 & 0 & h^{(3)}_{33} & 0 & 0 & \dots \\
    \ket{3,4} & & 0 & 0 & 0 & 0 & 0 & h^{(4)}_{33} & h^{(4)}_{35} & \dots \\
    \ket{5,4} & & 0 & 0 & 0 & 0 & 0 & h^{(4)}_{53} & h^{(4)}_{55} & \dots \\
    \vdots    & & \vdots & \vdots & \vdots & \vdots & \vdots & \vdots & \vdots & \ddots \\
    \end{block}
\end{blockarray}
\label{eq: Hamiltonian matrix}
\end{equation}
where $h^{(J)}_{NN'} = \braket{N, J|\mathcal{H}^{(\text{eff})}_v|N', J}$ are the non-zero matrix elements. The block-diagonal pattern seen in Eq.~\eqref{eq: Hamiltonian matrix} for $J \neq 0$ repeats for higher values of $J$. The $2 \times 2$ symmetric matrices corresponding to even values of $J \neq 0$ are easily diagonalized by an orthogonal transformation, and the eigenfunctions $\ket{F_\alpha,J}$ ($\alpha=1,2,3$) of the effective Hamiltonian can be written as linear combinations of the Hund's case ($b$) basis functions $\ket{N, J}$~\cite{larsson2019}
\begin{subequations}\label{eqs: eigenfunctions}
\begin{align}
    &\ket{F_1,J} = \cos\theta_J\ket{J-1,J} + \sin\theta_J\ket{J+1,J} \rm{,} \\
    &\ket{F_2,J} = \ket{J, J} \rm{,} \\
    &\ket{F_3,J} = -\sin\theta_J\ket{J-1,J} + \cos\theta_J\ket{J+1,J} \rm{,}
\end{align}
\end{subequations}
where the ``fine-structure labels'' $F_\alpha$ are introduced to distinguish the eigenfunctions corresponding to the same $J$~\cite{babcock1948}, and the mixing angles $\theta_J$ are given in terms of matrix elements in Eq.~\eqref{eq: Hamiltonian matrix} as
\begin{equation}
    \tan2\theta_J = \frac{2h^{(J)}_{J-1,\,J+1}}{h^{(J)}_{J-1,\,J-1}-h^{(J)}_{J+1,\,J+1}} \rm{.}
    \label{eq: mixing angle}
\end{equation}
Figure~\ref{fig: mixing angle} shows the mixing angles calculated using the effective Hamiltonian, Eq.~\eqref{eq: H_eff}, with the molecular constants taken from Yu \textit{et al.}~\cite{yu2014}. It can be seen that in the pure Hund's case ($b$) limit (i.e., for $J \gg 1$)
\begin{subequations}\label{eqs: case (b) limit}
\begin{align}
    &\ket{F_1,J} \longrightarrow \ket{J-1,J} \rm{,} \\
    &\ket{F_2,J} \longrightarrow \ket{J, J} \rm{,} \\
    &\ket{F_3,J} \longrightarrow \ket{J+1,J}.
\end{align}
\end{subequations}
The calculated mixing angles can be found in Supplementary Material.

Since certain \textit{E2} transitions can be accounted for only within the intermediate coupling approach, the limiting behavior expressed by Eqs.~\eqref{eqs: case (b) limit} allows for a qualitative comparison between the fine-structure resolved rotational transitions realizable in the pure Hund's case ($b$) and in the intermediate coupling \big(defined by Eqs.~\eqref{eqs: eigenfunctions}\big). 

\begin{figure}[h!]
    \centering
        \includegraphics[width=\linewidth]{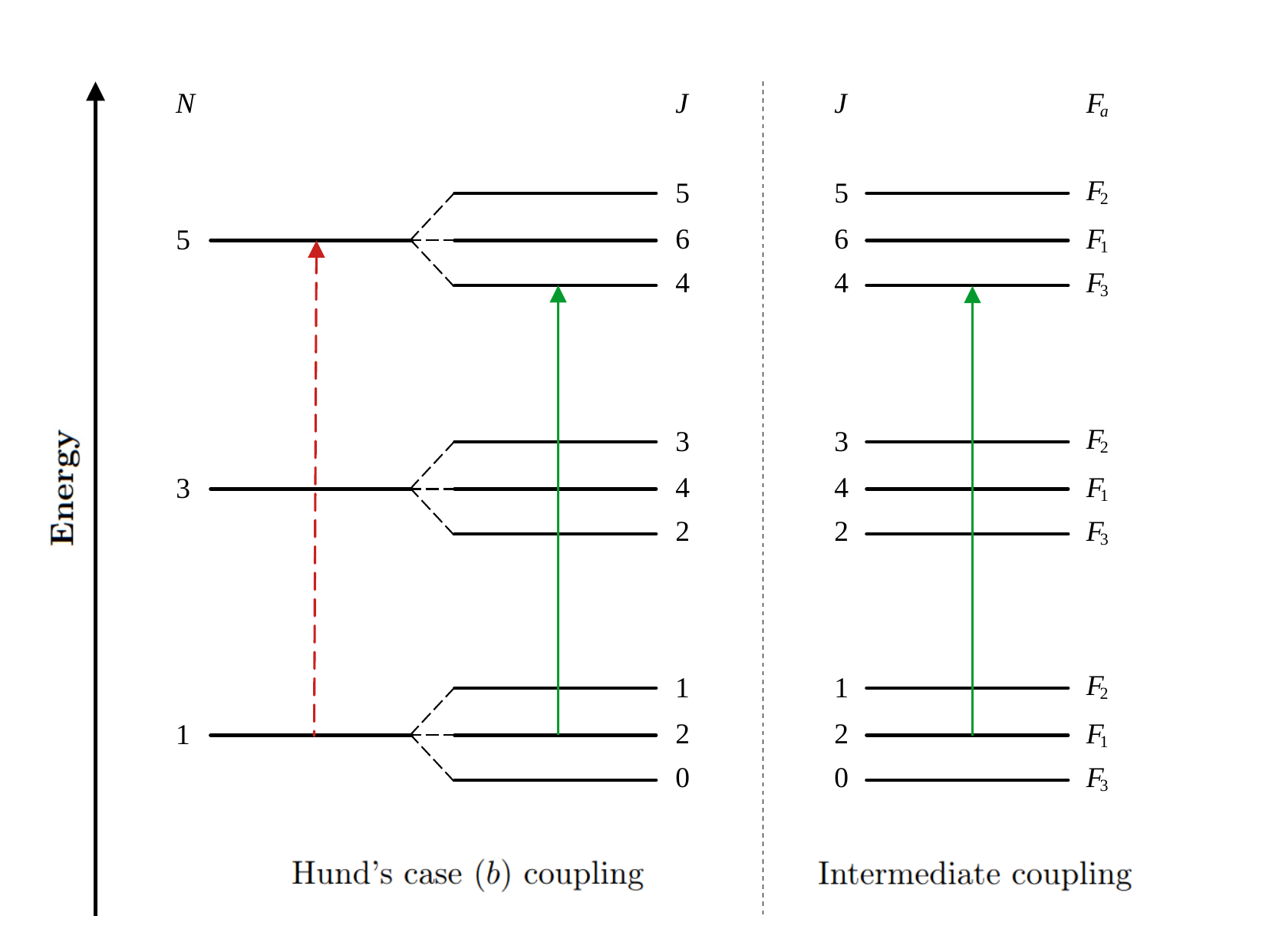}
    \caption{Schematic depiction (not to scale) of the fine-structure resolved rotational energy levels in $^{16}\text{O}_2(X^3\Sigma^-_g)$ obtained in the pure Hund's case ($b$) coupling (left) and in the intermediate coupling (right). In the intermediate coupling, the $F_3 \leftarrow F_1$ $S(2)$ transition is \textit{E2}-allowed (green solid arrow), while its Hund's case ($b$) counterpart -- denoted appropriately as $U(1)\,\,S(2)$ -- violates the additional $\Delta N = 0, \pm 2$ case ($b$) selection rules (red dashed arrow).}
    \label{fig: energy levels}
\end{figure}

An example of such transition is shown schematically in Fig.~\ref{fig: energy levels}. In the intermediate coupling, the $F_3 \leftarrow F_1$ $S(2)$ absorption line \big(here $S(2)$ corresponds to selection rules for the good quantum number $J$\big) is \textit{E2}-allowed, while in Hund's case ($b$) the corresponding $U(1)\,\,S(2)$ transition violates the $\Delta N = 0, \pm 2$ selection rules (see the Wigner symbols in Eqs.~\eqref{eq: Q orbital} \& \eqref{eq: L&L formula} below); the ordering in the $U(1)\,\,S(2)$ notation corresponds to $\Delta N \,\, \Delta J$ selection rules. Thus, the intensity of this transition cannot be calculated in pure Hund's case ($b$).

On the other hand, Eqs.~\eqref{eqs: case (b) limit} allow for retaining the $\Delta N \,\, \Delta J$ notation for labeling the fine-structure resolved rotational branches, even if they are obtained using the intermediate coupling, rather than pure Hund's case ($b$). In this case, the two notations: $M(5) \,\, O(4)$ and $F_1 \leftarrow F_3$ $O(4)$ refer to the same transition \big(cf. Eqs.~\eqref{eqs: case (b) limit}\big) -- provided that \textit{all} the discussed results correspond to the intermediate coupling. In particular, the \textit{E2} absorption lines in the oxygen $1$-$0$ fundamental band currently available in HITRAN~\cite{hitran2022} correspond to the intermediate coupling case, but the transitions themselves are labeled using the $\Delta N \,\, \Delta J$ notation. In this work, we perform all calculations using the intermediate coupling and we label the obtained fine-structure resolved transitions using both of the aforementioned notations (in the main text and in Supplementary Material). 

\subsection{Fine-structure resolved \textit{E2} spectral line intensities}
The total rovibronic wave function of $^{16}$O$_2$($X^{3}\Sigma^{-}_{g}$) can be written symbolically as $\ket{\Lambda\,S, v, F_\alpha, J M}$, where the electronic part\footnote{The calculations in this work are performed under the Born-Oppenheimer approximation.} $\ket{\Lambda\,S}$ is labeled with the total electronic spin $S$ and the component $\Lambda$ of the total electronic orbital angular momentum along the internuclear axis; the rovibrational part $\ket{v}=\ket{v(N)}$ depends jointly on the vibrational ($v$) and pure-rotational ($N$) quantum numbers due to the anharmonicity of the underlying molecular potential energy curve; and the spin-dependent rotational part $\ket{F_\alpha, J M}$ ($\alpha = 1, 2, 3$) is given by the intermediate-coupling eigenfunctions of Eq.~\eqref{eq: H_eff} that were defined in Eqs.~\eqref{eqs: eigenfunctions}. In the following, the constant value of $S=1$ will be omitted from the state-vector labels; the same will be true for the angular momentum space-fixed projection quantum number, $M$, which is just being summed-over in the line-intensity formula.

The intensity $S^{\text{FS}}_{fi}$ of the fine-structure resolved \textit{E2} absorption line, corresponding to a transition between the initial state $\ket{i} = \ket{\Lambda_i, v_i, F_i, J_i}$ and the final state $\ket{f} = \ket{\Lambda_f, v_f, F_f, J_f}$, is given (in SI units) by the expression~\cite{jozwiak2020}
\begin{equation}
    S^{\text{FS}}_{fi} = \frac{2\pi^{4}}{15hc^3\varepsilon_{0}} \frac{p_{i}(T)}{(2J_{i}+1)} \left(1 - e^{-h\nu_{fi}/k_{\rm B}T}\right) \nu_{fi}^{3} \left|\mathcal{M}^{\text{FS}}_{fi}\right|^{2} \rm{,}
    \label{eq: S_fi}
\end{equation}
in which $h$, $c$, $\varepsilon_0$ and $k_{\rm B}$ are the Planck constant, the vacuum speed of light, the vacuum permittivity and the Boltzmann constant, respectively; $p_i(T) = p_{v_i F_i J_i}(T)$ is the Boltzmann distribution of molecular energy levels $E_i = E_{v_i F_i J_i}$ at temperature $T$
\begin{equation}
    p_{v_i F_i J_i}(T) = \frac{(2J_i+1)e^{-E_{v_i F_i J_i}/k_{\rm B}T}}{Z(T)} \rm{,}
    \label{eq: Gibbs}
\end{equation}
with the partition function given by
\begin{equation}
    Z(T) = \sum\limits_{v, F_\alpha, J} (2J + 1) e^{-E_{v F_\alpha J}/k_{\rm B}T} \rm{.}
\end{equation}
$\nu_{fi} = (E_f-E_i)/h$ is the transition frequency and $\mathcal{M}^{\text{FS}}_{fi}$ is the spin-dependent \textit{E2} transition moment, whose squared modulus in Eq.~\eqref{eq: S_fi} is
\begin{align}
    \left|\mathcal{M}^{\text{FS}}_{fi}\right|^2 = \sum_{\substack{M_f, M_i,\\p}} &\left|\braket{\Lambda_f, v_f, F_f, J_f M_f|\mathsf{Q}^{(2)}_{\,\,p}|\Lambda_i, v_i, F_i, J_i M_i}\right|^2 \notag\\
    = &\left|\braket{\Lambda_f, v_f, F_f, J_f||\mathsf{Q}^{(2)}||\Lambda_i, v_i, F_i, J_i}\right|^2 \rm{,}
    \label{eq: |M^FS|^2}
\end{align}
where $\mathsf{Q}^{(2)}_{\,\,p}$ are the spherical tensor components~\cite{judd1975} of the \textit{E2} moment defined in the laboratory frame. The summation in Eq.~\eqref{eq: |M^FS|^2} is performed over the space-fixed components of the total angular momentum ($M_i$, $M_f$) and the \textit{E2} spherical tensor ($p$).

The \textit{E2} moment operator does not act on the spin degrees of freedom of the system. Hence the reduced matrix elements in Eq.~\eqref{eq: |M^FS|^2} can be written as~\cite{judd1975}
\begin{align}
    &\braket{\Lambda_f, v_f, F_f ,J_f||\mathsf{Q}^{(2)}||\Lambda_i, v_i, F_i, J_i} = \notag\\
    &(-1)^{J_i+S}[J_f, J_i]^\frac{1}{2}
    \sum_{N_f, N_i} C^{F_f}_{N_f} C^{F_i}_{N_i} (-1)^{N_f} \sixj{N_f}{2}{N_i}{J_i}{S}{J_f} \times
    \notag\\
    &\qquad\qquad\qquad\qquad\qquad \times \braket{\Lambda_f, v_f, N_f||\mathsf{Q}^{(2)}||\Lambda_i, v_i, N_i} \rm{,}
    \label{eq: Q orbital}
\end{align}
where $C^{F_\alpha}_{N_\alpha}$ are the mixing coefficients defined in Eqs.~\eqref{eqs: eigenfunctions}
\begin{subequations}\label{eqs: mixing coeffs}
\begin{align}
    &C^{F_1}_{J-1} = C^{F_3}_{J+1} = \cos\theta_J \rm{,}
    \\
    &C^{F_2}_{J} = 1 \rm{,}
    \\
    &C^{F_1}_{J+1} = -C^{F_3}_{J-1} = \sin\theta_J \rm{.}
\end{align}
\end{subequations}
The spin-independent reduced matrix element on the right-hand side of Eq.~\eqref{eq: Q orbital} is defined in the laboratory frame and can be related to the matrix element taken in the molecule-fixed frame by the formula~\cite{Landau}
\begin{align}
    \braket{\Lambda_f, v_f, N_f||\mathsf{Q}^{(2)}||\Lambda_i, v_i, N_i} = (-1)^{N_f-\Lambda_f} [N_f, N_i]^\frac{1}{2} \times \notag\\
    \times \threej{N_f}{2}{N_i}{-\Lambda_f}{\Delta\Lambda}{\Lambda_i} \braket{\Lambda_f, v_f(N_f)|Q^{(2)}_{\Delta\Lambda}|\Lambda_i, v_i(N_i)} \rm{,}
    \label{eq: L&L formula}
\end{align}
where $Q^{(2)}_{\,\,q}$ are the spherical tensor components of the \textit{E2} moment defined in the molecule-fixed frame and $\Delta\Lambda = \Lambda_f - \Lambda_i$. In the present case of the $X^3\Sigma^-_g \leftarrow X^3\Sigma^-_g$ transitions, $\Delta\Lambda = 0$ and we only need to consider matrix elements of the $Q^{(2)}_{\,\,0}$ component in Eq.~\eqref{eq: L&L formula}. In the molecule-fixed frame, with the origin at the center of mass and the $z$-axis taken along the internuclear axis, $Q^{(2)}_{\,\,0}$ coincides with the (ordinary) Cartesian $Q_{zz}$ component of the \textit{E2} moment tensor~\cite{buckingham1959}
\begin{equation}
    Q^{(2)}_{\,\,0} = \frac{1}{2}\sum_{n} e_n\left(3z_{n}^{2} - r^2_n\right) \rm{,}
    \label{eq: Q_zz}
\end{equation}
where $e_n$ is the $n$-th electric charge in the system (the summation includes both protons and electrons) whose radial distance from the origin is $r_n$.

Since in the Born-Oppenheimer approximation $\ket{\Lambda, v(N)} = \ket{\Lambda}\ket{v(N)}$, in Eq.~\eqref{eq: L&L formula} we first average $Q^{(2)}_{\,\,0}$ over the electronic state of O$_2$ and obtain the \textit{E2} moment curve
\begin{align}
    \mathcal{Q}(R) = \frac{|e|}{2}\left[ZR^2 - \braket{X^3\Sigma^-_g(R)|\sum_{\alpha}\left(3z^2_\alpha - r^2_\alpha\right)|X^3\Sigma^-_g(R)}\right] \rm{,}
    \label{eq: Q(R)}
\end{align}
where $R$ is the internuclear distance, $|e|$ is the value of the elementary electric charge, $Z$ is the atomic number of each of the identical nuclei, and the subsequent summation is taken over the electrons only. The functional form of Eq.~\eqref{eq: Q(R)} is obtained from Eq.~\eqref{eq: Q_zz} by noting that in a homonuclear molecule the center of mass coincides with half of the distance $R$ between the nuclei and that matrix elements of the nuclear part of Eq.~\eqref{eq: Q_zz} in the basis of electronic wave functions are trivial; note that the \textit{E2} moment curve depends on $R$ also through the parametric dependence of the electronic Born-Oppenheimer wave functions on $R$ (which is further indicated in Eq.~\eqref{eq: Q(R)}).

Thus, the desired matrix elements of $Q^{(2)}_{\,\,0}$ in Eq.~\eqref{eq: L&L formula} can be computed, with use of Eq.~\eqref{eq: Q(R)}, as
\begin{align}
    \mathcal{M}_{fi} = \int\limits_{0}^{\infty} \text{d}R\,\, \phi_{v_f, N_f}(R)\,\mathcal{Q}(R)\,\phi_{v_i, N_i}(R) \rm{,}
    \label{eq: M_fi}
\end{align}
where $\phi_{v, N}$ are the wave functions of the rovibrational states $\ket{v(N)}$, which can be made real-valued due to the problem being one dimensional~\cite{Landau} (note that the lower indices in $\mathcal{M}_{fi}$ do not correspond to the same quantum numbers as those in $\mathcal{M}^{\text{FS}}_{fi}$ -- the latter include the fine-structure quantum numbers, which is indicated by the superscript $\text{FS}$). Using Eqs.~\eqref{eq: |M^FS|^2}-\eqref{eq: L&L formula} together with Eq.~\eqref{eq: M_fi} allows for expressing the squared modulus of the spin-dependent transition moment $\mathcal{M}^{\text{FS}}_{fi}$ (defined in the laboratory frame) in terms of the spin-free transition moment $\mathcal{M}_{fi}$ (defined in the molecule-fixed frame) as follows:
\begin{align}
    |\mathcal{M}^{\text{FS}}_{fi}|^2 = [J_f, J_i]
    \Bigg|\sum_{N_{f},N_{i}} C^{F_f}_{N_f}\,C^{F_i}_{N_i}& [N_f, N_i]^\frac{1}{2} \times \notag\\ 
    \times \threej{N_{f}}{2}{N_{i}}{0}{0}{0}& \sixj{N_{f}}{2}{N_i}{J_{i}}{S}{J_f} \mathcal{M}_{fi}\Bigg|^2
    \rm{.}
    \label{eq: M_fi^FS}
\end{align}

\section{\label{Sec3}\textit{E2} moment curve, rovibrational wave functions and transition moments}
The \textit{E2} moment curve of $^{16}\text{O}_2(X^3\Sigma^-_g)$, Eq.~\eqref{eq: Q(R)}, was calculated using the 2022.2 version of the MOLPRO package~\cite{MOLPRO_brief:2022.2}, employing the MRCI method with the uncontracted double-augmented six-$\zeta$ quality (d-aug-cc-pV6Z) basis set~\cite{dunning:89a}. All electrons (including the core electrons) were distributed among the 10 lowest molecular orbitals. The calculations were performed for interatomic distances in the 0.70–3.50~{\AA} range, with a constant step of 0.01~{\AA}. The \textit{ab initio} \textit{E2} moment curve can be found in Supplementary Material. Based on numerical tests employing different basis sets, we estimate its accuracy at $\sim$10\%.

\begin{figure}[h!]
    \centering
    \includegraphics[width=\linewidth]{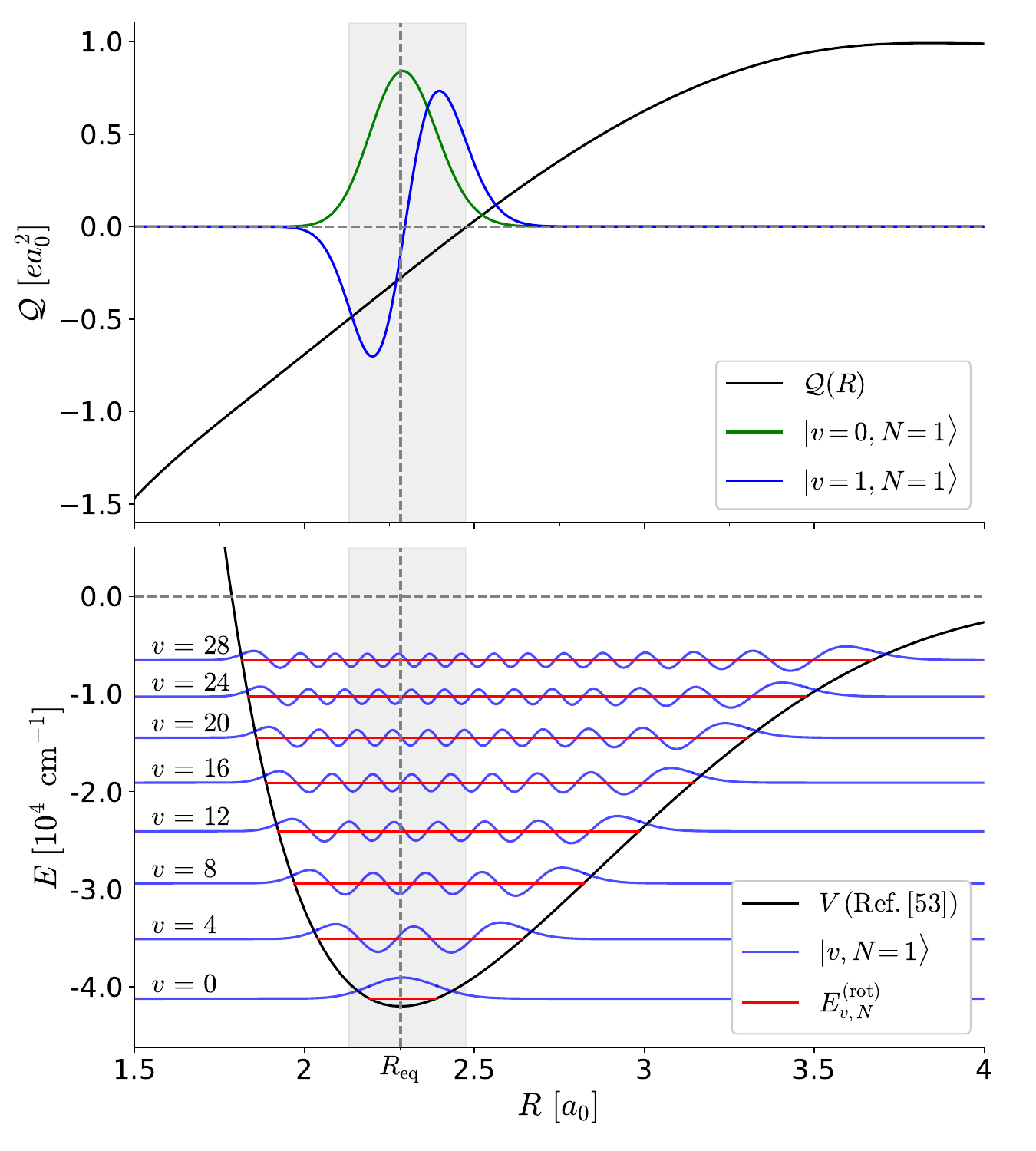}
    \caption{(upper panel) The \textit{ab initio} \textit{E2} moment curve (black line), Eq.~\eqref{eq: Q(R)}, and two sample rovibrational wave functions (green and blue lines). (lower panel) The \textit{ab initio} potential energy curve of $^{16}\mathrm{O}_2(X^{3}\Sigma^{-}_{g})$~\cite{bytautas2010} (black line) and sample rovibrational energy levels (red lines) with their corresponding wave functions (blue lines). The range of the internuclear distances covered by the experiments is indicated with a gray vertical strip around the equilibrium bond-length $R_{\text{eq}}$ (gray vertical dashed line). The gray horizontal dashed lines on both panels indicate the zero of the ordinates.}
    \label{fig: Q and V}
\end{figure}

The rovibrational wave functions, required for evaluating the spin-free transition moment integrals in Eq.~\eqref{eq: M_fi}, were calculated with the DVR method~\cite{lill1982, light1985} using the state-of-the-art \textit{ab initio} potential energy curve of $^{16}\text{O}_2(X^3\Sigma^-_g)$ reported by Bytautas, Matsunaga and Ruedenberg~\cite{bytautas2010}. $500$ basis functions were used in the calculations and the radial grid spanned $2000$ points between $0.5\,a_0$ and $10\,a_0$. Some of the rovibrational wave functions are shown in Fig.~\ref{fig: Q and V}. In the upper panel, the wave functions used for determining the $\braket{v_f=1|\mathcal{Q}|v_i=0}$ \textit{E2} transition moment of the fundamental band in O$_2$ are plotted against the \textit{ab initio} \textit{E2} moment curve. It can be seen that the bond-length dependence of the \textit{E2} moment must be taken into account when calculating the relevant transition moments, as in this case the integrals in Eq.~\eqref{eq: M_fi} cannot be reduced to a simple product of some permanent molecular multipole moment and the rovibrational Franck-Condon factor. In the lower panel, the rovibrational wave functions are shown together with the corresponding energy levels supported by the potential. The DVR wave functions can be found in Supplementary Material.

For the reasons which are given in \ref{appendix}, when calculating the spin-free transition moments using Eq.~\eqref{eq: M_fi} the \textit{E2} moment curve was represented by a rational function (Pad\'{e} function)
\begin{equation}
    \mathcal{Q}(R) = \frac{\sum\limits_{m=0}^{M} \alpha_m \left(\frac{R-R_0}{R_0}\right)^m}{\sum\limits_{n=0}^{N} \beta_n \left(\frac{R-R_0}{R_0}\right)^n} \rm{,}
    \label{eq: Pade}
\end{equation}
with $M=N=12$ and the value of $R_0 = 2.477\, a_0$ fixed (our \textit{E2} moment curve changes sign at $\sim$$R_0$). The coefficients $\alpha_m$ and $\beta_n$, determined from the least-squares fit to the \textit{ab initio} points, can be found in Supplementary Material. Due to the sufficiently large number of radial grid points for the DVR wave functions, the integrals in Eq.~\eqref{eq: M_fi} were calculated using the standard trapezoidal rule. All calculated spin-free transition moments, $\mathcal{M}_{fi}$, can be found in Supplementary Material. The spin-dependent transition moments squared, $|\mathcal{M}^{\text{FS}}_{fi}|^2$, can then be obtained using Eq.~\eqref{eq: M_fi^FS} with the mixing coefficients defined in Eqs.~\eqref{eqs: mixing coeffs}. For convenience, the calculated values of $|\mathcal{M}^{\text{FS}}_{fi}|^2$ are also included in Supplementary Material.

As was already mentioned in Introduction, there have been numerous studies, both empirical and \textit{ab initio}, devoted to the determination of the ``permanent'' \textit{E2} moment of $^{16}\text{O}_2(X^3\Sigma^-_g)$, understood either as the vibrational average over the bond-length dependence of the \textit{E2} moment or as its particular value taken, e.g., at the equilibrium internuclear O-O distance. These values are compiled in Tab.~\ref{tab: quadrupole moments 1}. As far as the \textit{E2} transition moments are concerned, it appears that the only available study is the one by Reid \textit{et al.}~\cite{reid1981}, in which the value of $0.108 \pm 0.019\,\, ea_0^2$ was empirically deduced for the $v = 1 \leftarrow 0$ vibrational transition moment. Table~\ref{tab: quadrupole moments 2} contains the vibrationally averaged (over the ground $v=0$ state) and the $v = 1 \leftarrow 0$ transition \textit{E2} moments calculated using Eq.~\eqref{eq: M_fi}. In both cases the lowest rotational $N=1$ state is assumed. In particular, the value obtained for the $v = 1 \leftarrow 0$ \textit{E2} transition moment agrees within the experimental uncertainty with the value obtained by Reid \textit{et al.}~\cite{reid1981}.

\begin{table}[h!]
    \centering
    \caption{Experimental and theoretical values of the ``permanent'' ground-state \textit{E2} moment of $^{16}\text{O}_2(X^3\Sigma^-_g)$ obtained throughout the years. The theoretical values were reported either as vibrational average of the \textit{E2} moment tensor or as the value of the \textit{ab initio} \textit{E2} moment function for a particular internuclear distance $R$. In Refs.~\cite{birnbaum1976, cohen1977} only the absolute value of the \textit{E2} moment was obtained.}
    \begin{tabular}{l|S[table-format=2.5]}
        \toprule
        {Ref. (year)} & {value $[ea_0^2]$} \\
        \midrule
        \cite{buckingham1968} (1968) & -0.3 \\
        \midrule
        \cite{birnbaum1976} (1976) &  0.26 \\
        \midrule
        \cite{cohen1977} (1977) & 0.25 \\
        \midrule
        \cite{becke1982} (1982) & -0.36 \\
        \midrule
        \cite{vanLenthe1984} (1984) & -0.264 \\
        \midrule
        \cite{visser1985} (1985) & -0.3146 \\
        \midrule
        \cite{laaksonen1985} (1985) & -0.3892 \\
        \midrule
        \cite{rijks1989} (1989) & -0.27137 \\
        \midrule
        \cite{hettema1994} (1994) & -0.264063 \\
        \midrule
        \cite{bundgen1995} (1995) & -0.249 \\
        \midrule
        \cite{dickinson1996} (1996) & -0.356 \\
        \midrule
        \cite{lawson1997} (1997) & -0.2273 \\
        \midrule
        \cite{minaev2004} (2004) & -0.254 \\
        \midrule
        \cite{bartolomei2010} (2010) & -0.2251 \\
        \midrule
        \cite{couling2014} (2014) & -0.230 \\
        \midrule
        \cite{kalugina2015} (2015) & -0.21 \\
        \midrule
        \cite{somogyi2024} (2024) & -0.223 \\
        \bottomrule
    \end{tabular}
    \label{tab: quadrupole moments 1}
\end{table}

\begin{table}[h!]
    \centering
    \caption{Matrix elements $\braket{v_f,N_f|\mathcal{Q}|v_i,N_i}$ calculated using Eq.~\eqref{eq: M_fi}. The result obtained for the $v = 1 \leftarrow 0$ \textit{E2} transition moment agrees with the empirically determined value $0.108 \pm 0.019\,\, ea_0^2$~\cite{reid1981}.}
    \begin{tabular}{c|S[table-format=2.5]}
        \toprule
        {matrix element} & {value $[ea_0^2]$} \\
        \midrule
        $\braket{0, 1|\mathcal{Q}|0, 1}$ & -0.26416 \\
        $\braket{1, 1|\mathcal{Q}|0, 1}$ &  0.09989 \\
        \bottomrule
    \end{tabular}
    \label{tab: quadrupole moments 2}
\end{table}

\section{Line intensities, Einstein \textit{A}-coefficients and comparison with HITRAN2020}
\begin{figure}[h!]
    \centering
    \includegraphics[width=\linewidth]{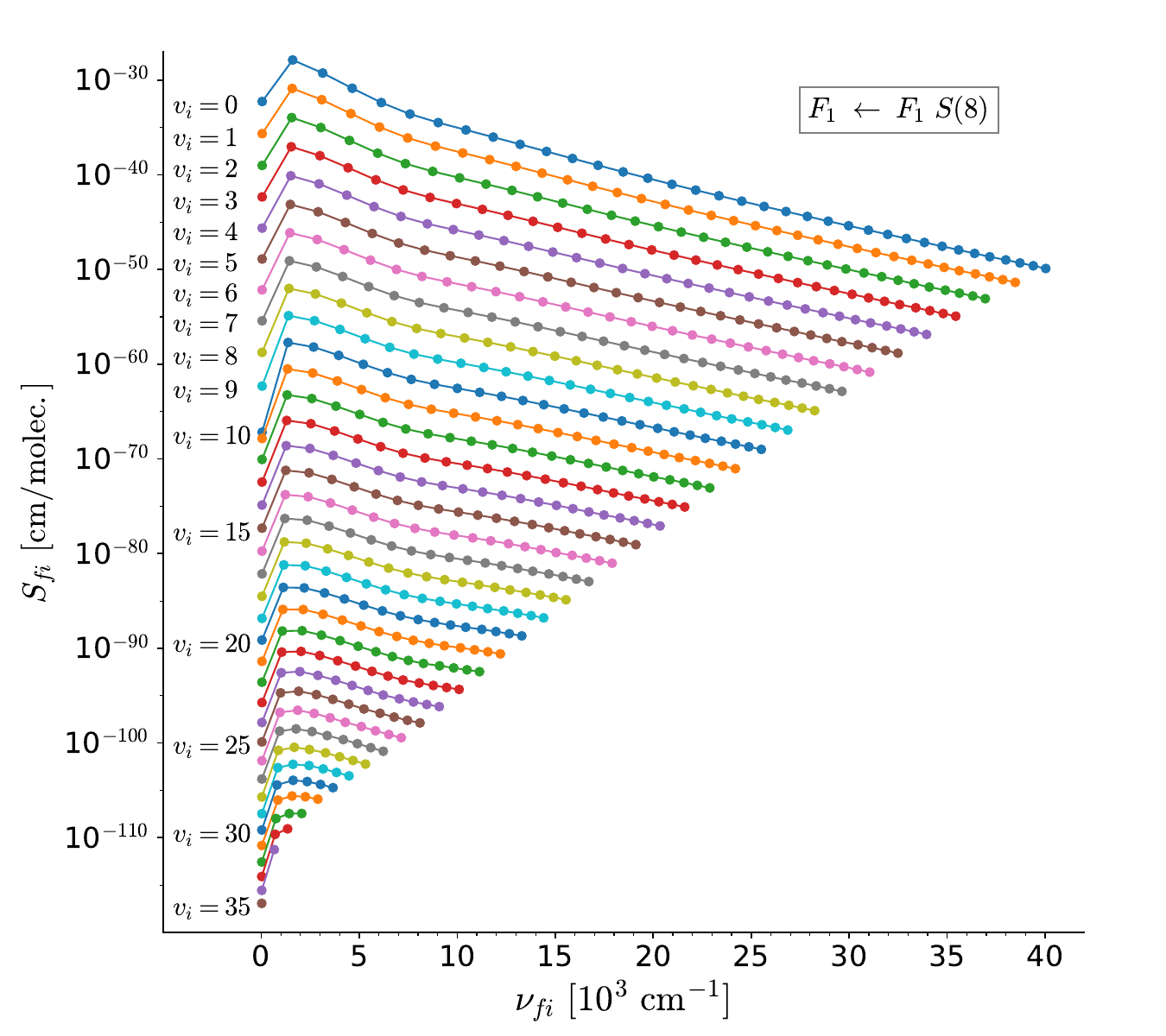}
    \caption{The intensity of the $F_1 \leftarrow F_1$ $S(8)$ line at $T_0 = 296$~K, shown as a function of the absorption frequency for all vibrational bands with $v \leq 35$. The different colored curves correspond to common initial vibrational quantum number $v_i$ (for clarity, only some values of $v_i$ are indicated explicitly). The rightmost point of each individual colored curve corresponds to the final vibrational quantum number $v_f=35$.}
    \label{fig: all intensities}
\end{figure}

We consider all fine-structure resolved rovibrational \textit{E2} absorption lines in $^{16}\text{O}_2(X^3\Sigma^-_g)$ for which the vibrational quantum number is $v \leq 35$ and the total angular momentum quantum number is $J \leq 40$. This corresponds to $666$ vibrational bands, both overtone and hot bands, comprising a total number of $280\,188$ individual spectral lines. The intensities of these lines were calculated using Eq.~\eqref{eq: S_fi} for the temperature $T_0=296$~K. The relevant transition frequencies $\nu_{fi}$ and the Boltzmann probabilities $p_i(T_0)$ were determined using the energy levels and partition function $Z(T_0)=215.7364$ reported by Yu \textit{et al.}~\cite{yu2014}. The SI unit of the intensity $S^{\text{FS}}_{fi}$ in Eq.~\eqref{eq: S_fi} is m$^{2}$/(s $\times$ molecule). To allow for a comparison with HITRAN, $S^{\text{FS}}_{fi}$ must be further divided by the vacuum speed of light $c$ and brought to the unit of cm/molecule \big(note that this is equivalent to cm$^{-1}$/(molecule $\times$ cm$^{-2}$)\big)~\cite{simeckova2006}. The resulting intensities are denoted by $S_{fi}$, so that we have
\begin{equation}
    S_{fi} = \frac{100}{c} \times S^{\text{FS}}_{fi} \rm{.}
    \label{eq: unit conversion}
\end{equation}
In this work, we report the calculated absorption intensities as $S_{fi}$. Figure~\ref{fig: all intensities} shows a sample frequency dependence of the $F_1 \leftarrow F_1$ $S(8)$ transition, which is the strongest calculated absorption line at $T_0$, for all the vibrational overtones and hot bands. 

For convenience, we have also calculated the temperature-independent Einstein \textit{A}-coefficients (in the usual SI units of 1/s)~\cite{Penner}
\begin{equation}
    A^{\text{FS}}_{fi} = \frac{16 \pi^5}{15 h c^5 \varepsilon_0} \frac{1}{2J_f+1} \nu_{fi}^5 \left|\mathcal{M}^{\text{FS}}_{fi}\right|^{2}
\end{equation}
corresponding to the expression in Eq.~\eqref{eq: S_fi}. The value of $S^{\text{FS}}_{fi}$ for an arbitrary temperature $T$ can then be obtained as
\begin{equation}
    S^{\text{FS}}_{fi} = \frac{c^2}{8 \pi \nu_{fi}^2} \frac{2J_f+1}{2J_i+1} p_i(T) \left(1 - e^{-h\nu_{fi}/k_{\rm B}T}\right) A^{\text{FS}}_{fi}
\end{equation}
and brought to the HITRAN-adopted units using Eq.~\eqref{eq: unit conversion}. All the calculated line intensities $S_{fi}$ and \textit{A}-coefficients $A^{\text{FS}}_{fi}$ can be found in Supplementary Material.

In Fig.~\ref{fig: six-fig}, we compare our calculated absorption intensities with those available in the latest 2020 edition of the HITRAN database~\cite{hitran2022}. It was already mentioned in Introduction that as far as the $X^3\Sigma^-_g \leftarrow X^3\Sigma^-_g$ rovibrational \textit{E2} transitions in $^{16}\text{O}_2$ are concerned, the data presently available in HITRAN includes only the $1$-$0$ fundamental band. Figure~\ref{fig: six-fig} shows the calculated \textit{E2} absorption intensities for the five allowed fine-structure resolved $J$-branches in the $1$-$0$ vibrational band. The \textit{ab initio} intensities for the transitions which are also available in HITRAN are indicated with color-filled points, while those not included in the database are marked using color-edged white points. The ``vs. HITRAN'' residuals in Fig.~\ref{fig: six-fig} given for each $J$-branch are obtained as absolute values of relative differences (in percent) between our \textit{ab initio} and the HITRAN intensities, relative to the HITRAN value. It can be seen that all the calculated \textit{E2} absorption intensities for the $1$-$0$ band differ from those available in HITRAN at the $5\% - 12\%$ level. Taking into account the $\sim$10\% accuracy of our \textit{ab initio} \textit{E2} moment curve, we therefore recommend using the intensities of the vibrational overtones and hot bands calculated in this work for updating the HITRAN database in the new 2024 edition.

\begin{figure*}
    \centering
    \includegraphics[width=\linewidth]{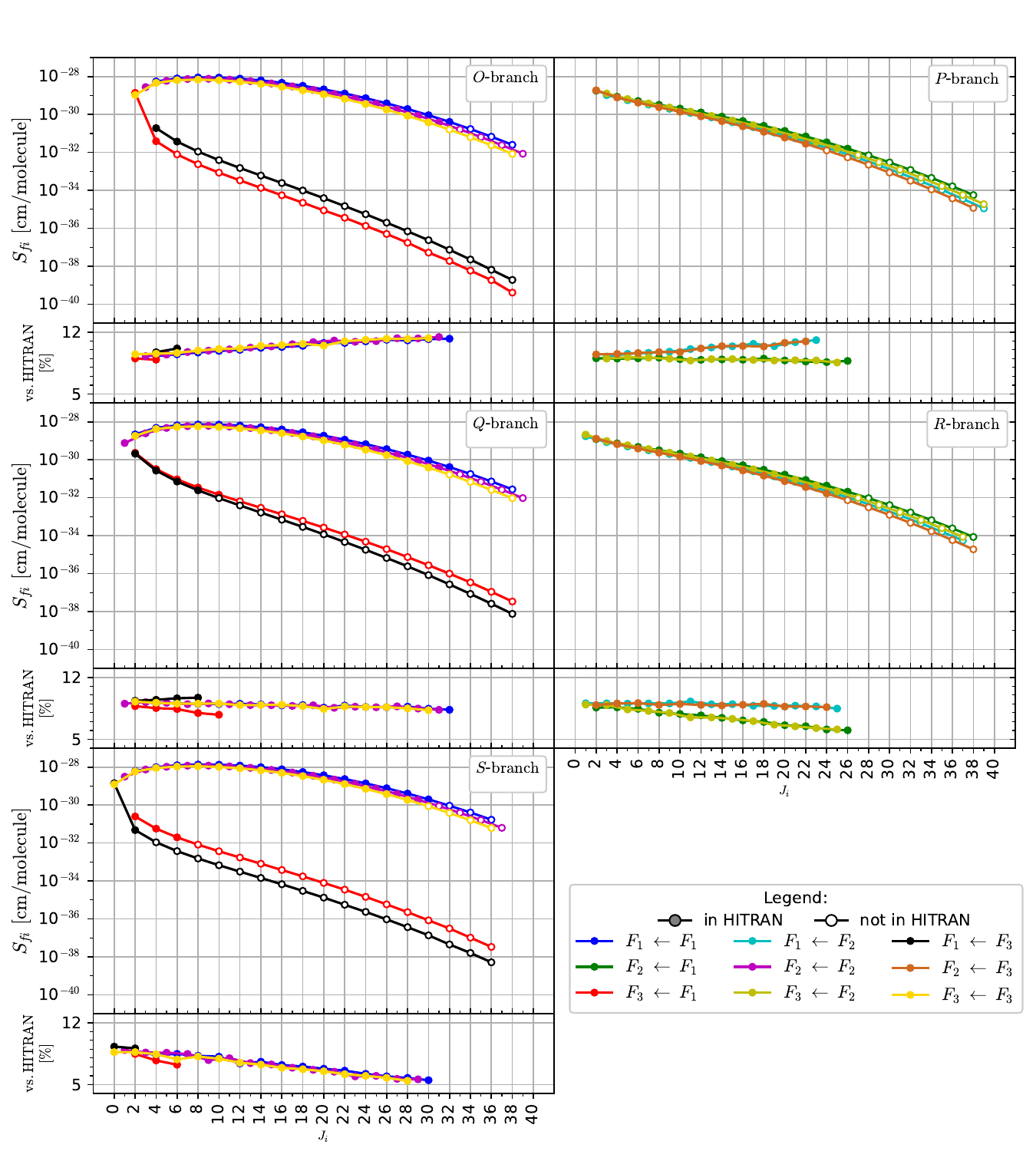}
    \caption{Comparison between our \textit{ab initio} and HITRAN intensities of the \textit{E2} absorption lines in the $1$-$0$ fundamental band of $^{16}\text{O}_2(X^3\Sigma^-_g)$. The figure consists of five panels, one for each indicated \textit{E2}-allowed \textit{J}-branch. In each panel, the \textit{ab initio} intensities are shown in the upper part and the absolute values of relative differences (in percent) between the \textit{ab initio} and HITRAN values are shown in the lower part. If a transition is available in HITRAN2020~\cite{hitran2022}, it is indicated with a color-filled point; if not, it is marked with a color-edged white circle. The accuracy of the \textit{ab initio} \textit{E2} moment curve used in obtaining the relevant absorption intensities is $\sim$10\%.}
    \label{fig: six-fig}
\end{figure*}

\begin{figure*}
    \centering
    \includegraphics[width=\linewidth]{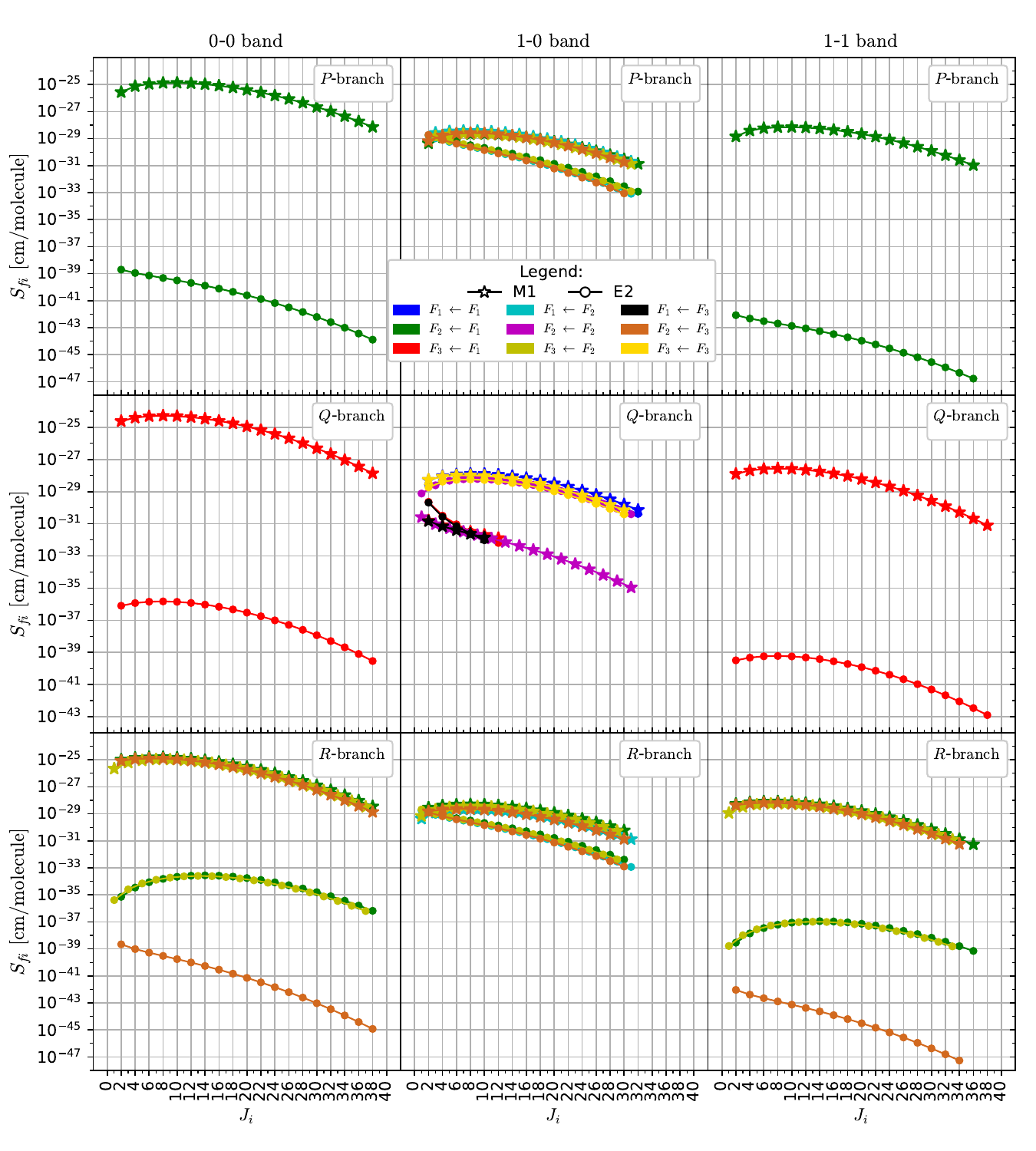}
    \caption{Comparison between the \textit{M1} and \textit{E2} absorption intensities in the $0$-$0$, $1$-$0$ and $1$-$1$ vibrational bands of $^{16}\text{O}_2(X^3\Sigma^-_g)$. The \textit{M1} intensities (indicated with stars) are taken from HITRAN, while the corresponding \textit{E2} intensities (marked with points) are calculated in this work (HITRAN2020~\cite{hitran2022} includes the \textit{E2} absorption lines only for the $1$-$0$ band, see Fig.~\ref{fig: six-fig}). The figure consists of three columns and three rows. Each column corresponds to a different indicated vibrational band and each subsequent row corresponds to a different \textit{M1}- and \textit{E2}- allowed \textit{J}-branch.}
    \label{fig: six-fig-2}
\end{figure*}

Besides the $1$-$0$ fundamental band, HITRAN contains the data on the \textit{M1} absorption intensities in the $0$-$0$ and $1$-$1$ pure rotational bands of the electronic $X^3\Sigma^-_g \leftarrow X^3\Sigma^-_g$ transition in $^{16}\text{O}_2$. Having calculated the intensities of the relevant rovibrational \textit{E2} absorption lines, it is interesting to compare the relative strength of the \textit{M1} and \textit{E2} components comprising the $0$-$0$, $1$-$0$ and $1$-$1$ vibrational bands, which might also prove useful in refining the available \textit{M1} intensities (see Introduction). Such a comparison is shown in Fig.~\ref{fig: six-fig-2}, whose three columns correspond to the three indicated vibrational bands and the three subsequent rows correspond to the three possible \textit{J}-branches resulting from the common \textit{M1} and \textit{E2} selection rules. The relevant \textit{M1} intensities taken from HITRAN~\cite{hitran2022} are marked with stars and the \textit{E2} intensities calculated in this work are indicated with points.

\section{Conclusion}
We have considered all fine-structure resolved rovibrational electric quadrupole (\textit{E2}) absorption lines in $^{16}\text{O}_2(X^3\Sigma^-_g)$ corresponding to the vibrational quantum numbers $v \leq 35$ and the total angular momentum quantum numbers $J \leq 40$. We calculated a new \textit{ab initio} \textit{E2} moment curve of O$_2$ and used it for obtaining the relevant \textit{E2} transition moments. We computed the temperature-independent Einstein \textit{A}-coefficients and the absorption intensities at $T=296$~K for all the considered transitions. We performed all the calculations in the intermediate coupling, i.e., using the exact eigenvectors of the effective Hamiltonian of O$_2$, which includes the relativistic spin-spin interaction-induced mixing of rotational states of O$_2$. The agreement between our \textit{ab initio} intensities and those available in HITRAN~\cite{hitran2022} \big(which at present includes only the $1$-$0$ fundamental band of $^{16}\text{O}_2(X^3\Sigma^-_g)$\big) prompts us to recommend incorporating the data on the vibrational overtones and hot bands reported here in the new 2024 edition of the HITRAN database. 

\subsection*{Note added during manuscript revision}
In a recently published work~\cite{somogyi2024}, Somogyi, Yurchenko and Kim calculated various electronic characteristics of the $X^3\Sigma^-_g$, $a^1\Delta_g$, $b^1\Sigma^+_g$, $I^1\Pi_g$, $II^1\Pi_g$, $I^3\Pi_g$ and $II^3\Pi_g$ terms of $^{16}\text{O}_2$, including the electric quadrupole (\textit{E2}) moment curve of the $X^3\Sigma^-_g$ term perturbed to first order by the spin-orbit coupling (SOC) to the excited $b^1\Sigma^+_g$, $I^1\Pi_g$ and $II^1\Pi_g$ states. The corresponding mixing coefficients have absolute values of the order $\sim 10^{-2}$, $\sim 10^{-4}$ and $\sim 10^{-3}$, respectively~\cite{somogyi2024, klotz1985}. Since the overall contribution of the SOC effects to the line intensity in Eq.~\eqref{eq: S_fi} is much less than the accuracy of the MRCI calculations, we neglect them in the present work. 

In contrast to Ref.~\cite{somogyi2024}, in our MRCI calculations all the electrons were distributed among the $10$ lowest molecular orbitals and a large, uncontracted, double-augmented basis set of six-$\zeta$ quality was used. This allowed us to calculate the values of the \textit{E2} moment for all the considered bond lengths, including the most important region (approximately covered by the shaded areas in Fig.~\ref{fig: Q and V}) around the equilibrium bond length, $R_{\rm eq}$.

\section*{Acknowledgments}
We thank Ioulli Gordon for bringing our attention to the issue of unphysical saturation of the high-overtone transitions and for his suggestions regarding the comparison of our \textit{E2} data with the HITRAN \textit{M1} data. M.G. was supported by the Polish Ministry of Science and Higher Education through Project No. PN/01/0229/2022 under the ``Per{\l}y Nauki'' program. H.J. was supported by the National Science Centre in Poland through Project No. 2019/35/B/ST2/01118. P.W. was supported by the National Science Centre in Poland through Project No. 2022/46/E/ST2/00282. For the purpose of Open Access, the authors has applied a CC-BY public copyright licence to any Author Accepted Manuscript (AAM) version arising from this submission. We gratefully acknowledge Polish high-performance computing infrastructure PLGrid (HPC Center: ACK Cyfronet AGH) for providing computer facilities and support within computational grant no. PLG/2024/017376. Created using resources provided by Wroclaw Centre for Networking and Supercomputing (\url{http://wcss.pl}).

\appendix
\section{\label{appendix}Analytical representation of the \textit{E2} moment curve}
The observation (made in Sec.~\ref{Sec3}) that the \textit{E2} moment cannot be taken outside the integral in Eq.~\eqref{eq: M_fi} introduces the problem of analytical representation of $\mathcal{Q}(R)$, as the computation of the relevant transition moments (especially for higher vibrational overtones and hot bands, for which the integrand oscillates rapidly) requires interpolating or fitting the \textit{ab initio} points with a continuous function. Since, in general, the appropriate functional form of the multipole moment (i.e., its bond-length dependence) is not known for a particular molecular potential, it would seem reasonable to simply interpolate the \textit{ab initio} points using splines of given order. It turns out, however, that this straightforward approach leads to spurious results and causes the calculated transition moments to saturate (as a function of the final-state vibrational quantum number) around a constant value in an unphysical way. To remedy this, the bond-length dependence of the multipole moment should be represented by functions which are analytic (in the sense of complex variable) on the real axis, rather than with the non-analytic spline interpolation.

The simplest explanation for this is obtained by noting that for high vibrational overtones the initial- and final-state energies are vastly different (cf. the lower panel of Fig.~\ref{fig: Q and V}) and, consequently, the product of the associated wave functions $\phi_{v_i,N_i}(R)$ and $\phi_{v_f,N_f}(R)$ in Eq.~\eqref{eq: M_fi} is rapidly oscillating, while $\mathcal{Q}(R)$ plays the role of a slowly-varying ``envelope'' for these oscillations (cf. the upper panel of Fig.~\ref{fig: Q and V}). In this case, it is well known~\cite{Zeldovich, Migdal} that each point $R$ at which either $\mathcal{Q}(R)$ or its derivatives (of given order) are discontinuous contributes to the integral by altering its value with respect to the one obtained when $\mathcal{Q}(R)$ is a smooth, analytic function. Since the interpolating splines are defined piecewise between the interpolation knots, their derivatives (in the case of cubic splines, the third-order ones) are discontinuous at these knots. Thus, with spline interpolation every \textit{ab initio} point of the \textit{E2} moment curve becomes a singularity of the integrand in Eq.~\eqref{eq: M_fi}, which results in a constant artificial contribution to $\mathcal{M}_{fi}$ from all discontinuity points of the spline derivatives.

If $\mathcal{Q}(R)$ is analytic on the real axis (like the Pad\'{e} function in Eq.~\eqref{eq: Pade}) then the qualitative behavior of the transition moments $\mathcal{M}_{fi}$ for high vibrational overtones -- i.e., for rapidly oscillating integrands in Eq.~\eqref{eq: M_fi} -- can be obtained using a quasi-classical approximation~\cite{Landau, Migdal}, the form of this dependence being determined by the underlying molecular potential curve. In particular, Medvedev~\cite{medvedev1985} proposed to consider overtone transitions as a quasi-classical tunneling problem (example of the so-called dynamical tunneling~\cite{heller1999, medvedev2012}) and used the Landau-Lifshitz formula for quasi-classical matrix elements~\cite{Landau} to estimate the high transition-frequency dependence of $\mathcal{M}_{fi}$ under the assumption that the repulsive short-distance part of the molecular potential can be approximated using an exponential function with a singularity at $-\infty$ (as is the case with, e.g., the familiar Morse potential). The result (given as Problem $1$ in \S51 of Ref.~\cite{Landau}) is then that the logarithm of the modulus of $\mathcal{M}_{fi}$ is proportional to the square root of the transition frequency\footnote{Strictly speaking, it is proportional to the difference between the square roots of the energies of the states between which a transition occurs (cf. Problem $1$ in \S51 of Ref.~\cite{Landau}). In the case of overtone transitions, however, the initial-state energy is fixed and can be gauged as a zero of the energy scale, as is done in this work. The final-state energy under the remaining square root becomes a transition frequency in this case.} $\nu_{fi}$. This result, termed the ``natural intensity distribution law'' (NIDL)~\cite{medvedev1985}, can be used to monitor the asymptotic behavior of the calculated values of $\mathcal{M}_{fi}$ to ensure that no unphysical ``non-analytic'' features set in at higher vibrational overtones.

\begin{figure}[h!]
    \centering
    \includegraphics[width=\linewidth]{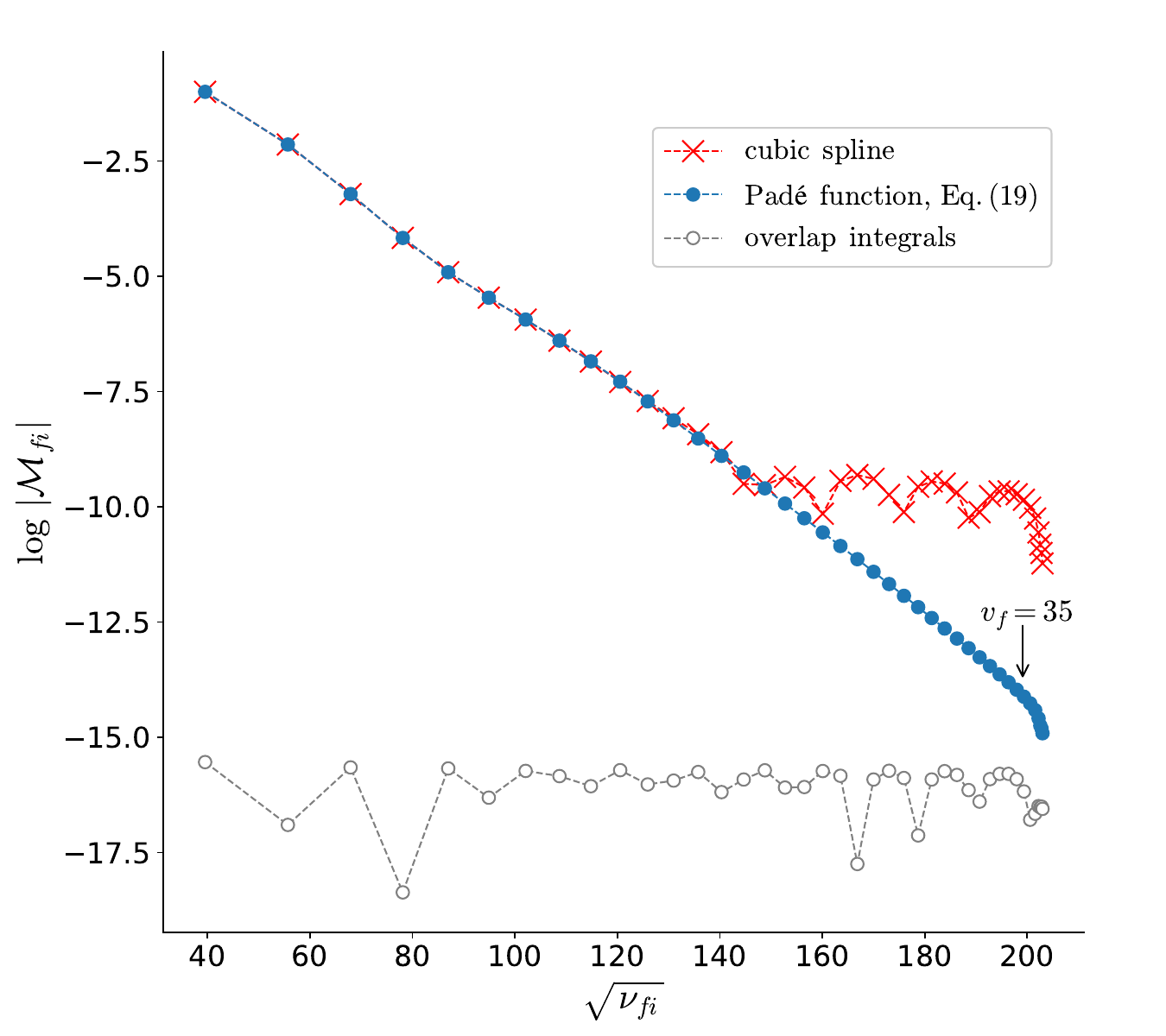}
    \caption{Logarithm of the modulus of the spin-free transition moment $\mathcal{M}_{fi}$, Eq.~\eqref{eq: M_fi}, as a function of the square root of the transition frequency, $\nu_{fi}$, between the fixed initial state $\ket{i}=\ket{v_i=0,N_i=1}$ and the final states $\ket{f}=\ket{1 \leq v_f \leq 41,N_f=1}$. Interpolating the \textit{E2} moment curve using piecewise-defined splines results in an unphysical saturation of $\mathcal{M}_{fi}$ for high overtones (red crosses). This effect does not appear when the \textit{E2} moment is fitted using a continuously-defined Pad\'{e} function, Eq.~\eqref{eq: Pade}, analytic on the real axis (blue points). The DVR wave functions are calculated using quadruple precision to ensure that the overlap integrals $\braket{f|i}$ (grey points) fulfill the orthogonality condition with sufficient numerical accuracy.  The highest measured~\cite{yang1989} vibrational level $v_f=35$ is indicated with an arrow.}
    \label{fig: saturation}
\end{figure}

Figure~\ref{fig: saturation} shows a comparison between the transition moments $\mathcal{M}_{fi}$ calculated using the spline-interpolated and the Pad\'{e}-fitted \textit{E2} moment curves, plotted using the ``log vs. square root'' NIDL variables~\cite{medvedev1985}. The aforementioned unphysical saturation effect, associated with the non-analytic character of the spline interpolation, sets in at $v_f \sim 15$ and is clearly visible. In contrast, the Pad\'{e} function in Eq.~\eqref{eq: Pade} gives a uniform high-frequency dependence of $\mathcal{M}_{fi}$, in accord with the qualitative NIDL prediction. The slight downwards bend of the blue points in Fig.~\ref{fig: saturation} is associated with approaching the dissociation limit~\cite{medvedev2015}, which in the ground-electronic state of O$_2$ occurs at $v_f \sim 41$. In this work, we consider all rovibrational transitions up to (and including) the $v_f = 35$ level, indicated in Fig.~\ref{fig: saturation} with an arrow, which is the highest measured vibrational level in O$_2$~\cite{yang1989}. To further ensure that the analytical form of the \textit{E2} moment curve is the only factor limiting the accuracy of the calculated values of $\mathcal{M}_{fi}$, we also computed the overlap integrals between the relevant rovibrational wave functions and plotted them using the same NIDL variables in Fig.~\ref{fig: saturation}. We found that quadruple precision is required for the calculated DVR wave functions to fulfill the orthogonality conditions with sufficient accuracy -- if this was not the case, the gray points in Fig.~\ref{fig: saturation} would ``move upwards'' on the plot and result in a similar artificial saturation in the high-frequency ``tail'' of the blue points. This observation is in accord with the findings of previous studies done for the electric dipole transitions in carbon monoxide~\cite{medvedev2015, medvedev2016, meshkov2018, meshkov2022, medvedev2022} and phosphorus nitride~\cite{ushakov2023}.

\begin{figure}[h!]
    \centering
    \includegraphics[width=\linewidth]{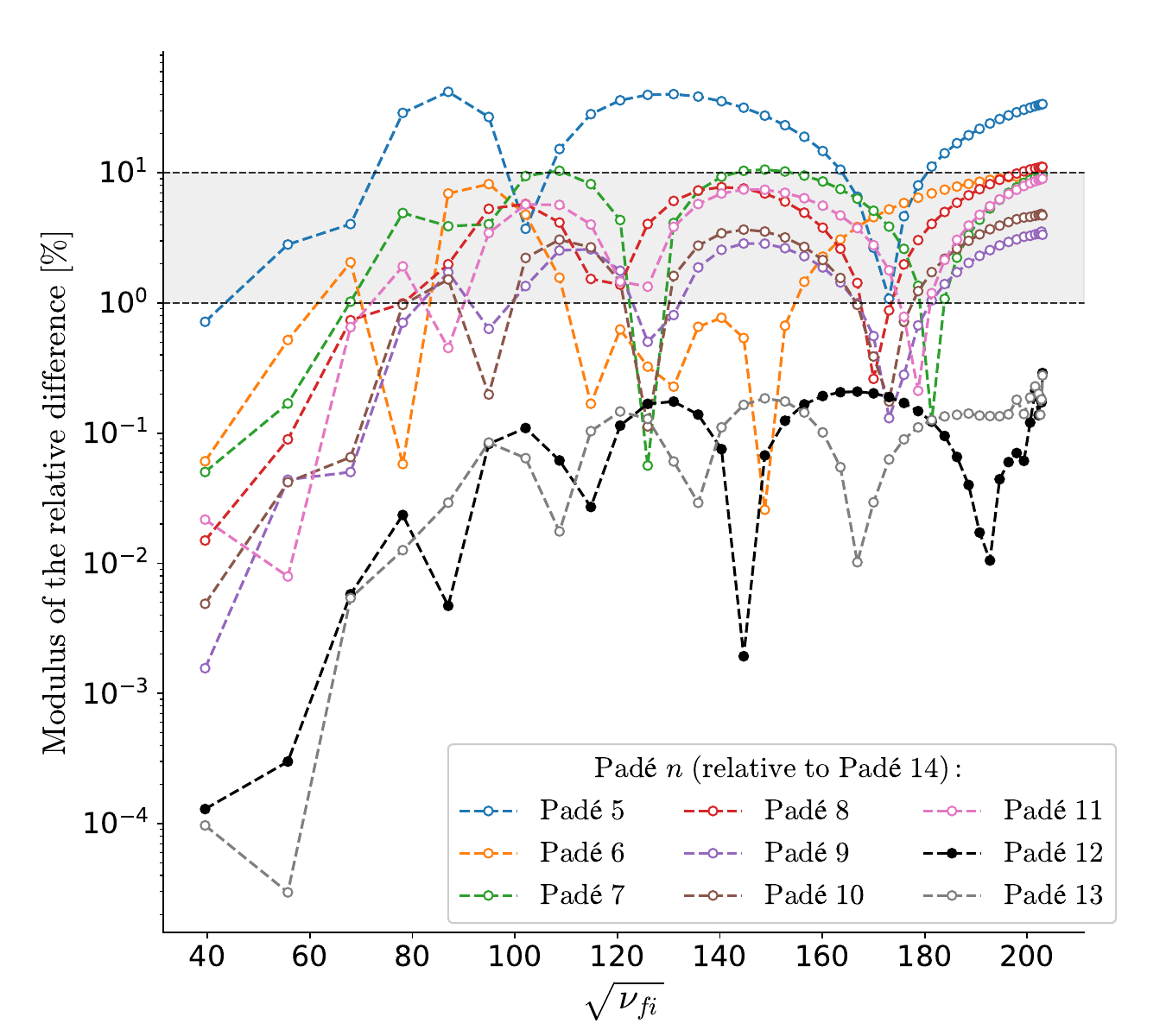}
    \caption{Modulus of the relative difference (in percent) between the values of $|\mathcal{M}_{fi}|^2$ calculated using $n^{\text{th}}$-degree Pad{\'e} functions specified in the plot legend, relative to the one obtained with $14^{\text{th}}$-degree Pad{\'e} function. The abscissa and the rovibrational transitions considered are the same as in Fig.~\ref{fig: saturation}. The gray horizontal strip covers the region between $1\%$ and $10\%$, the latter being the estimated accuracy of our \textit{ab initio} \textit{E2} moment curve.}
    \label{fig: pade accuracy}
\end{figure}

The Pad\'{e} function in Eq.~\eqref{eq: Pade} was chosen as the simplest rational function\footnote{It is known~\cite{king1979} that rational functions, rather than polynomial expansions, tend to better represent the fitted data in regions far from the expansion center.} for which the bond-length dependence of $\mathcal{Q}(R)$ in Eq.~\eqref{eq: M_fi} is analytic on the real axis. The values of $M=N=12$ in Eq.~\eqref{eq: Pade} were chosen as a compromise between the number of the fitting parameters and the internal consistency of the calculated line intensities. This is illustrated in Fig.~\ref{fig: pade accuracy}, which shows the absolute value of the relative difference between the squared modulus $|\mathcal{M}_{fi}|^2$ of the \textit{E2} transition moment (to which the line intensity is directly proportional) calculated using ``Pad{\'e}~$n$'' functions (i.e., using Eq.~\eqref{eq: Pade} with $M=N=n$) relative to the results obtained with ``Pad{\'e} $14$'' function. It can be seen that the line intensities calculated using Eq.~\eqref{eq: Pade} with $M=N=12$ differ on the sub-percent level relative to those obtained with the ``Pad{\'e} 14'' function for all overtone transitions.

Of course, due to the aforementioned lack of explicit expression for the bond-length dependence of molecular multipole moments, the choice of the analytic function fitted to the \textit{ab initio} points is not unique. At best, one may constrain the possible forms of these functions, e.g., by investigating the qualitative asymptotic behavior of the multipole moment of a diatomic molecule in the united- and the separated-atom limits. Actually, such asymptotic behavior was found in the case of the molecular electric dipole moment~\cite{goodisman1963}, which allowed for constraining the forms of the analytic functions used in the treatment of the electric dipole transitions in hydrogen halides: HF, HCl, HBr and HI~\cite{ogilvie1980, li2013}, in CO~\cite{medvedev2016, meshkov2018, medvedev2021, meshkov2022, medvedev2022} and in PN~\cite{ushakov2023}. It appears, however, that no such analysis have been done so far for the \textit{E2} moment.

\section*{Supplementary material}
Supplementary material associated with this article can be found in the online RepOD repository at doi: \url{https://doi.org/10.18150/JZDWVT}

\bibliographystyle{elsarticle-num}
\bibliography{references}

\end{document}